\documentstyle[aps,prd,graphics,epsfig]{revtex}

\def\zp {Z^{\prime}}
\def\vs {\vspace{0.0cm}}

\def\physl#1#2#3{Phys. Lett. B#1, #2 (#3)}

\def\nucl#1#2#3{Nucl. Phys. B#1, #2 (#3)}

\begin{document}
\twocolumn[\hsize\textwidth\columnwidth\hsize\csname
@twocolumnfalse\endcsname 

\title{More evidence in favour of Light Dark Matter particles?}
\author{C\'eline B\oe hm$^1$, Yago Ascasibar$^2$}
\address{$^1$Astrophysics department, 1 Keble Road, OX1 3RH, Oxford, England, UK;\\
$^2$Harvard-Smithsonian Center for Astrophysics, 60 Garden St., Cambridge, MA 02138;\\
boehm@astro.ox.ac.uk, yago@head.cfa.harvard.edu}
\date{\today} 
\maketitle

\begin{abstract}
In a previous work, it was found that the Light Dark Matter (LDM)  scenario could be a possible explanation to the 511 keV emission line 
detected at the centre of our galaxy. 
Here, we show that hints of this scenario may also 
have been discovered in particle physics experiments. This could 
explain the discrepancy between the measurement of the fine structure constant and the value referenced in the CODATA. 
Finally, our results indicate that some of the LDM features could be tested in accelerators. 
Their discovery might favour $N=2$ supersymmetry.
\end{abstract}

\vs
\vs

]

\section*{Introduction}
Despite many experimental and theoretical efforts, the origin of almost 95$\%$ of the universe remains 
unknown. About one third of the content of the Universe seems to be made of matter but only a few 
percent is to attribute to ordinary matter. The rest, called 
Dark Matter (DM), is expected to be made of neutral weakly-interacting massive particles. 
Their mass is generally believed to be above the proton mass.
However, recent studies have pointed out that lighter particles are possible, and perhaps even more promising 
from the astrophysical point of view.

\vs
Light Dark Matter particles are supposed to annihilate into pairs of leptons-anti leptons. 
When they lose their energy and decay (or annihilate, as in the case of 
electrons and positrons), they produce energetic photons that can be detected by modern gamma-ray telescopes.

\vs
The total DM annihilation cross-section  can be written as 
$ \sigma v_r \equiv a + b v^2.$ In this expression, $v_r$ and $v$ are the DM relative and individual velocity, 
respectively. In the case of the LDM scenario, this cross-section must satisfy two constraints in order to be 
compatible with current observations:
Firstly, it must have the correct value to explain why Dark Matter constitutes precisely 23$\%$ of the universe 
\cite{wmap}.
This value is about
$(\sigma v_r)_{\vert prim} \sim 10^{-26} \ \mbox{cm}^3\ \mbox{s}^{-1}$, and it is valid in the early universe (when $v\sim c/3$).
Secondly, it must decrease with time: $(\sigma v_r)_{MW}$ in the Milky Way must be $\sim 10^{-5}$ times smaller than 
$(\sigma v_r)_{\vert prim}$ to avoid an overproduction of low-energy gamma rays within our galaxy \cite{bens}. 

\vs
The combination of these two constraints yields
$a \lesssim 10^{-31} (m_{dm}/\rm{MeV})^2$ cm$^3$ s$^{-1}$ and $b \sim 10^{-25} (m_{dm}/\rm{MeV})^2$  cm$^3$ s$^{-1}$, 
valid at any time. 
%
From the particle physics point of view, the problem is to find a candidate with the appropriate  annihilation cross section.
A simple solution consists in coupling DM particles to a new neutral gauge 
boson ($\zp$). In this case, there is no a-term ($a=0$) and the b-term can be set to the correct value by imposing specific 
values of the $\zp$ couplings \cite{bf}.
Strictly speaking, DM could also be coupled to new particles $F$ having a mass $m_F > 100$ GeV (to satisfy accelerator limits). 
The net effect is to introduce an a-term in $\sigma v_r$, which must be smaller than 
$10^{-31} (m_{dm}/\rm{MeV})^2$ cm$^3$ s$^{-1}$ to satisfy the gamma-ray constraint \cite{bf}. 

\vs
The LDM scenario has been shown to provide an elegant explanation of the 511 keV emission line detected at the center of our galaxy.
Many experiments had measured the intensity of this emission over 
the years, but they lacked the resolution to determine its morphology.
Recently, however, INTEGRAL/SPI experiment 
provided a surface brightness map which unambiguously indicates the presence of 
an extended source located at the galactic center \cite{integral}. The observed flux is in good agreement with previous 
data \cite{previous}, and the morphology is well reproduced by a 2D-Gaussian with a Full Width Half Maximum (FWHM) of $\sim 10$ deg.

\vs
This gamma-ray emission line can be interpreted in terms of $e^+ e^-$ annihilations, although the
origin of low-energy positrons remains a matter of heated debate. Standard explanations (e.g. 
astrophysical sources) have been proposed, but most of them seem excluded by the value of the 
bulge-to-disc ratio \cite{pjean} or rely on rather 
controversial hypotheses (jet emission \cite{Prantzos}, positron propagation \cite{Bertone}, etc.). 
In contrast, the LDM scenario seems to be a quite appealing and simple explanation despite it certainly 
involves more exotic physics \cite{511}. In this framework, the DM particles (with a mass $m_{dm} < m_{\mu}$) would annihilate into electron-positron pairs.
Assuming a microgauss-scale magnetic field, the latter remain confined in the galactic center,
losing all their kinetic energy by collisions with the baryonic material and eventually annihilating into monoenergetic photons of 511 keV.

Here, we shall see that the LDM scenario must have two important ingredients. Both should be of experimental/theoretical interest.

\section{Can the morphology of the emission be well reproduced by LDM annihilations?}
\begin{figure}[h] 
\centering{\includegraphics[width=8cm]{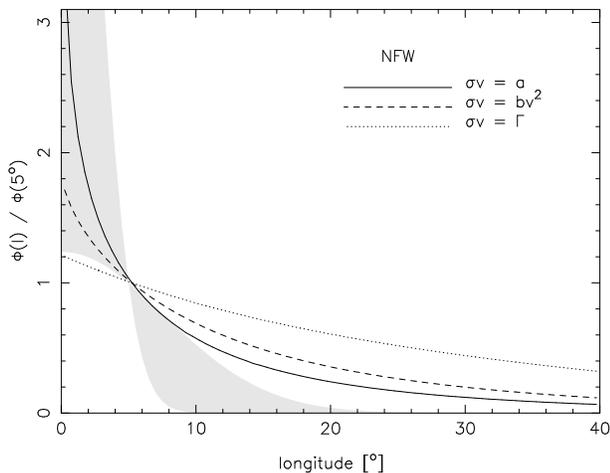}}
\caption{
Galactic 511 keV emission, integrated over $\pm15\deg$ in latitude and normalized to the flux at $l=5\deg$. 
Shaded area encloses approximate 2-$\sigma$ limits from INTEGRAL/SPI. Solid and dashed lines show the predictions 
for LDM with $b=0$ and $a=0$ respectively. 
Dotted line represents the emission profile expected for decaying dark matter. 
A NFW model has been used for the Milky Way halo. } \label{figSPI} 
\end{figure}

The answer is given in Figure~\ref{figSPI}, where we plotted the predictions of the LDM model, compared to the INTEGRAL/SPI results. 
We used a NFW profile to describe the dark matter halo of the Milky way, and a Gaussian source with FWHM 
between 6 and $18\deg$ (reported 2-$\sigma$ confidence limits) to fit the observational data.
Assuming any reasonable mass distribution for our galaxy (i.e. $\rho\sim r^{-3}$ at large radii) and 
using a realistic description of its velocity dispersion profile, we obtain that 
LDM annihilations produce a significant amount of positrons, even far away from the galactic center. 
However, the 511 keV emission is expected to be mostly from the bulge, since the baryon density in the outer parts of the halo is simply too small for the positrons to lose their kinetic energy.
For a fixed total flux, one obtains less photons from the outer regions when a cuspy density profile is assumed, in better agreement with INTEGRAL/SPI. 
On the other hand, the dark matter profile cannot be too cuspy, because the possibility of a point-like source is excluded by the data with a high confidence level. A thorough analysis of which halo models might be compatible with the observed morphology of the 511 keV line emission will be presented elsewhere.

\vs
For a NFW profile \cite{nfw} with $\rho_s=0.183$ GeV\,cm$^{-3}$ and $r_s=25$\,kpc, the total flux at the Earth ($r_\odot=8.5$\,kpc) is
\begin{equation} 
\Phi_{tot}= m_{MeV}^{-2} ( 130\,a_{26}+ 1.37\!\times\!10^{-4}\,b_{26} )\ {\rm cm}^{-2}\,{\rm s}^{-1} 
\end{equation} 
that we expressed in terms of the dimensionless parameters 
$m_{MeV}\equiv m_{dm}/(1\,{\rm MeV})$, $a_{26}\equiv a/(10^{-26}\,{\rm cm}^{-3}\,{\rm s}^{-1})$ 
and $b_{26}\equiv b/(10^{-26}\,{\rm cm}^{-3}\,{\rm s}^{-1})$.
Due to our naive prescription for positron propagation (each $e^+$ is assumed to instantly yield two 511 keV photons), most of the emission comes from large
angular distances from the galactic center. A fairer comparison with SPI data can be obtained by restricting 
ourselves to the inner $16\deg$. This roughly corresponds to the instrument's field of view, and  encloses most of the 
detected emission. Using the NFW density profile, we now obtain
\begin{equation} 
\Phi_{16}= m_{MeV}^{-2} (46.3\,a_{26}+2.98\!\times\!10^{-5}\,b_{26})\ {\rm cm}^{-2}\,{\rm s}^{-1}.
\label{eqF}
\end{equation}
The surface brightness beyond $16 \deg$ is expected to be below
$m_{MeV}^{-2}(70a_{26}+7\!\times\!10^{-5}b_{26})\ {\rm cm}^{-2}\,{\rm s}^{-1}\,{\rm sr}^{-1}$,
or even smaller because of the low baryon density outside the galactic center.
This is likely to be too faint to be detected by SPI, so a NFW profile is able to fit the observed 511 keV emission if the total flux is equated to expression (\ref{eqF}).
A more precise assessment of the best-fit values of the coefficients $a_{26}$, $b_{26}$ and $m_{MeV}$ 
requires a realistic treatment of the SPI instrumental response matrix.

\vs
A preliminary analysis hinted that a ``pure'' velocity-dependent cross-section (i.e. $\zp$ exchange only) could be entirely responsible for the 511 keV signal. 
Here, we show that the emission profile arising from this term is shallower than the one due to the $a$-term. 
This is a generic conclusion, regardless of the particular model assumed for the Milky Way mass. Therefore, our present conclusion is that, 
unlike stated in previous studies, the $a$-term should be included indeed. Moreover, accounting for the observed flux (with a $b$-term only) would require 
$m_{dm}<1$ MeV, which would have a tremendous impact on primordial nucleosynthesis.
Setting $\Phi_{16}=10^{-3}\ {\rm cm}^{-2}\ {\rm s}^{-1}$ (the INTEGRAL's result), we conclude that $a_{26}$ must be of the order of
\begin{eqnarray} 
a_{26} &\sim& 2.16 \ 10^{-5} \ m_{MeV}^{2}  \label{eqMdm} \\
\mbox{or} \ \
(\sigma v_r)_{MW} &\sim&  2.16 \ 10^{-31} \ m_{MeV}^{2} \ \mbox{cm}^3 \ \mbox{s}^{-1} \nonumber
\end{eqnarray} 
in order to explain the 511 keV line, assuming that LDM annihilations constitute the 
main source of low-energy positrons.

\vs
Finally, we note that:
i) fermionic LDM candidates cannot account for the observed flux, because
$a_{26} \ll  10^{-7} \ m_{MeV}^2$ for these particles (except perhaps in 
the marginal case where they would exchange a light gauge boson with axial couplings to ordinary matter 
\cite{llbfs}; in this case $m_{dm}$ should be $\sim$ 10 MeV),  
ii) LDM annihilations should proceed through both F and $\zp$ exchanges.
The extra gauge boson is necessary to obtain the correct relic density, while the F exchange 
is needed to explain the the 511 keV signal. We confirm that decaying DM cannot fit the emission in a NFW profile \cite{Hooper}.

\vs 
The need for (fermionic) $F$ particles is of interest for both atomic and particle physics experiments.  
Their contribution to the muon and electron anomalous magnetic moment ($g-2)_{\mu,e}$ is 
given by $\delta a_{\mu, e} \sim \frac{f_l f_r}{16 \pi^2} \frac{m_{\mu, e}}{m_{F_{\mu, e}}}$. 
Noting that the quantity $\frac{f_l f_r}{m_{F_{e}}}$ also enters the expression of $\sigma v_r$, 
we obtain
$(\sigma v_r)_{MW} \simeq  (0.864, 3.456) \ 10^{-31} \left(\frac{\delta a_{e}}{10^{-12}}\right)^2 $ cm$^3$ s$^{-1}$ 
(depending on whether DM is made of self-conjugate particles or not). $(\sigma v_r)_{MW}$ is dominated by its a-term.
The above expression matches eq.\ref{eqMdm} when $\delta a_e = (a_e^{exp} - a_e^{th}) \simeq (1.58,0.79) \ 10^{-12} \ m_{MeV}$ (respectively). 
Yet, we expect the experimental value $a_e^{exp}$ to be somewhat larger than the 
theoretical estimate $a_e^{th}$.

\vs
It turns out that there exists a small discrepancy between the theoretical prediction and the  
experimental measurement. The latter is about $ \delta a_e \sim (3.44-3.49) \ 10^{-11}$ 
(the first number is obtained from the positron g-2, while the second one is from the electrons) \cite{Marciano}. 
This discrepancy is generally ``disregarded'' because the new physics processes that are generally considered 
are supposed to yield a much smaller contribution. Here, we see that DM particles with a mass of $m_{dm} \sim (21.8,43.6)$ MeV 
(depending on whether DM is made of self-conjugate particles or not) could surprisingly explain 
the discrepancy. On the other hand, greater masses would yield a too large value of $\delta a_e$.

\vs
The discrepancy between $a_e^{exp}$ and $a_e^{th}$ thus appears for the first time related to a new physics process. 
To determine $a_e^{th}$, we used the fine structure constant obtained from the Quantum Hall effect
$\alpha_{QH}$. The latter is seen to be the most accurate experimental determination of $\alpha$. However, it is never used by particle 
physicists, since it leads to an unexplained discrepancy. 
To obtain a perfect agreement between theory and observations, one instead ``forces'' $a_e^{th}$ to match 
$a_e^{exp}$.
The value of $\alpha$ thus obtained (denoted $\alpha_{st}$)
is quoted in the international reference CODATA \cite{Mohr},
and it is used to get theoretical estimates of other processes. 
But this procedure yields wrong results if the discrepancy between $a_e^{exp}$ and $a_e^{th}$ is due to new physics. 
If it is so, then one should use the experimental value of $\alpha$ (for example $\alpha_{QH}$) instead of $\alpha_{st}$ 
to make theoretical predictions. 

\vs
The difference between $\alpha_{st}^{-1}$ and $\alpha_{QH}^{-1}$ is about $ 4 \ 10^{-6}$. This changes $a_e^{th}$ and presumably also 
$a_{\mu}^{th}$ (at least slightly). The latter was found to be smaller than the experimental value by a few $10^{-9}$ units \cite{E821}. This discrepancy is widely 
considered as a possible case for new physics and its precise value is of crucial interest. So, it would be useful to determine 
$a_{\mu}^{th}$ by taking $\alpha = \alpha_{QH}$ and estimate the discrepancy again.

\vs
We based our previous estimate of $\delta a_e = a_e^{exp}-a_e^{th}$ on $\alpha_{QH}$. 
Other experimental values of $\alpha$ can be found in the literature 
(obtained notably from the measurement of the Rydberg constant, Josephson effect and muonium). The most precise values are thought to 
come from the Quantum Hall effect and the Rydberg constant, and they both give $a_e^{exp}-a_e^{th}>0$.
Nevertheless, if we use the other two values we obtain a negative discrepancy, 
which cannot be explained by the presence of F particles, at least not in the simple form considered here (although they are not excluded neither).

\vs
Assuming universality and $F$ exchange, 
we obtain a relationship between the electron and muon g-2:
\begin{equation}
 \left(\frac{\delta a_{\mu}}{10^{-9}}\right) = 2.1 \ (m_{Fe}/m_{F \mu}) \ \left(\frac{\delta a_{e}}{10^{-11}}\right).
 \label{eq1}
 \end{equation}
 Plugging the measured discrepancy for $\delta a_e$ into the above expression and assuming $m_{Fe} = m_{F \mu}$, 
 we  find  $\delta a_{\mu} \sim 7 \ 10^{-9}$.  The E821 experiment measured $\delta a_{\mu} \sim (2.7 \pm 1.04) \ 10^{-9}$ 
 (using $e^+ e^-$ data). One can therefore explain both the anomalous values of the muon and electron g-2 by introducing a set of 
 $F$ particles satisfying $m_{Fe} \sim m_{F \mu}/x$, with presumably $x \gtrsim 2$. 
 The correct approach to determine $x$ more precisely would be to estimate $a_{\mu}^{th}$ 
 by using $\alpha_{QH}$. But as mentioned earlier, this has not been done yet. 
 Note that, to our knowledge, it is the first time that a possible connexion between $\delta a_e$ and $\delta a_{\mu}$ is established.
 One could make a prediction for the tau $g-2$. 
 However present experiments still lack the sensitivity to challenge the Standard Model predictions. 

 \vs 
Unambiguous signatures of $F$ particles could be detected in the Large Hadron Collider (LHC), 
unless they turn out to be heavier than a few TeV. They could be produced through 
$e^+ e^-$ collisions and detected through their two-body decay (\textit{i.e.} $F$ going into electrons and DM particles). 
It is worthwhile to precise that $F$ particles are required to explain the 511 keV line.
If the LDM scenario is not the explanation to this emission, then the spin-1 boson becomes the only ingredient that is 
really needed for LDM to be viable.

\begin{figure}[h]
{\includegraphics[width=4.cm]{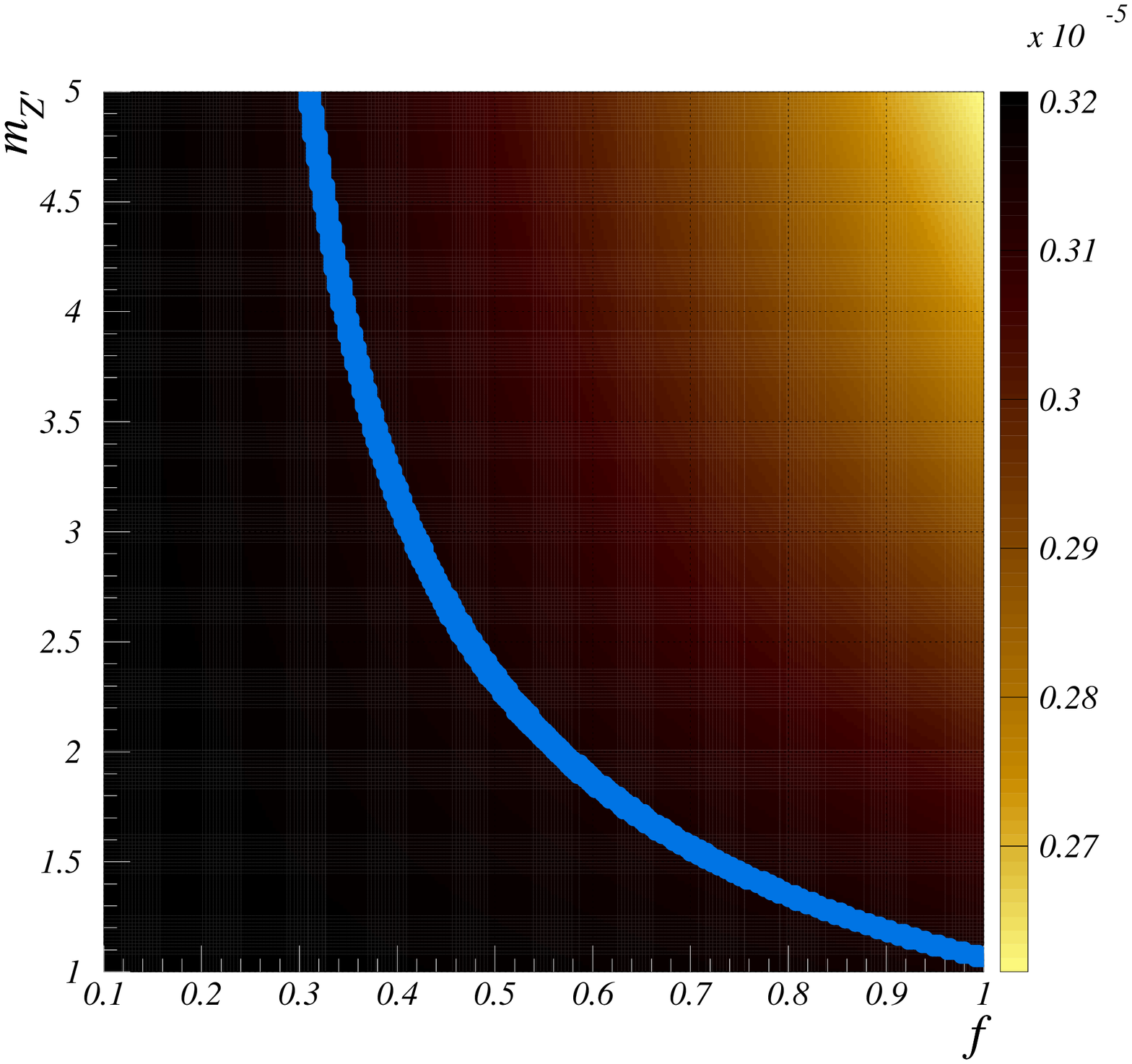} \includegraphics[width=4cm]{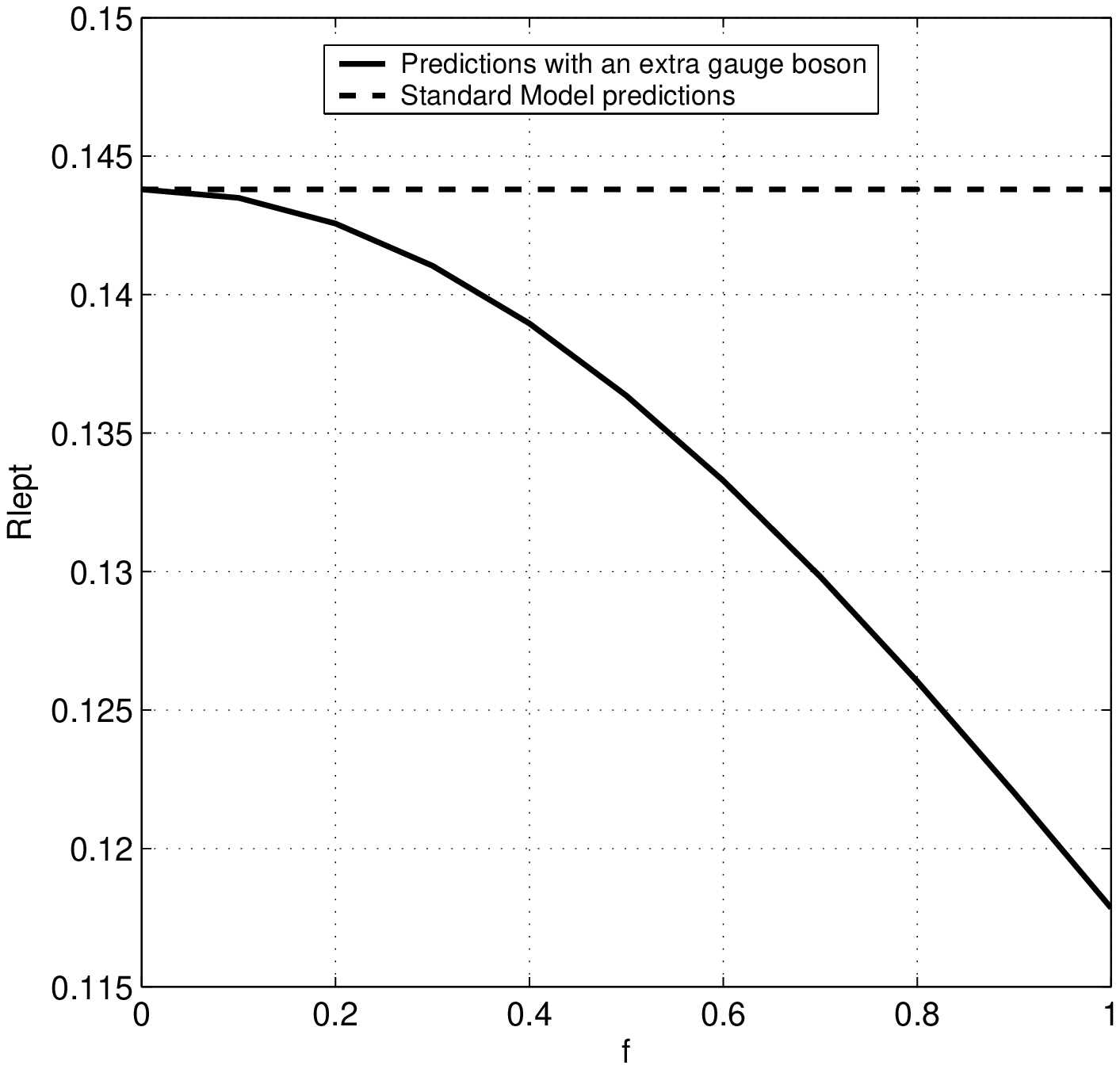}}
\caption{The left panel represents how NuTeV's observable $R_{num}$ 
is affected by a $\zp$. $R_{num}$ values in presence of a $\zp$ are symbolised by the colours ranging from black to yellow. 
The Standard Model predicts $R_{num}=   \sum_{u,d} \left[ G_F  c_v^q c_a^q  \right] = 3.2072 \ 10^{-6}$ (black region) while 
NuTeV finds $3.1507 \ 10^{-6}$. The parameters of the gauge boson that impressively fit the NuTeV anomaly (without error bars) are represented 
by the blue curve with $f_q f_{\nu}= f^2 \ 10^{-6} \ (m_{\zp}/\rm{GeV})^2$. On the right panel, we show our predictions for $R_{lept}$, using  
$f_e f_{\nu} = f^2 \ 10^{-6} \ (m_{\zp}/\rm{GeV})^2$.}
\label{nacio} 
\end{figure}

\vs
Hints of the presence of a light gauge boson may also have been  detected \cite{moi}.  
Assuming universal couplings, it was found that, albeit small, the $\zp$ couplings
could be ``large'' enough to modify the neutrino-quark elastic scattering cross sections when the Z' mass is about a few GeV.  
NuTeV collaboration measured these cross sections 
and found small deviations \cite{nutev}.  QCD corrections 
(isospin violations, strange sea asymmetry) could well be the explanation of this anomaly \cite{Kretzer}. 
However, QCD uncertainties are still very large and, in some cases,  even increase the anomaly. 
Therefore, it is worthwhile to consider the existence of a light gauge boson  seriously. 
Figure~\ref{nacio} illustrates how a light $\zp$ can solve the NuTeV anomaly.
(Note that a QCD explanation does not exclude the existence of a $\zp$, but it certainly sets
an upper limit on its mass.)

\vs
 The case without universality seems even more interesting as one could perhaps find evidence for a light $\zp$ in high energy colliders! 
The measurement of the electron $g-2$ restricts the coupling $f_e$ to be less than  
$f_e \lesssim 5.7 \ 10^{-1} \  (\frac{\delta a_{e}}{10^{-11}})^{1/2} \ (\frac{m_{\zp}}{\rm{GeV}})$. 
Also, the product of the $\zp$ couplings to neutrinos and electrons must not exceed
$[f_e f_{\nu}]^{max} \sim 5.388 \, 10^{-7} \ (\frac{m_{\zp}}{\rm{GeV}})^2,$ 
according to the very precise measurement of the elastic scattering of muon neutrinos on electrons by the CHARM II experiment \cite{charmii}.
CHARM II's results are in good agreement with the Standard Model (SM) predictions. However, the mean experimental value seems to be slightly 
different from the SM expectations. It is quite surprising but very interesting to see that 
the introduction of a new gauge boson with couplings $f_e f_{\nu} \simeq [f_e f_{\nu}]^{max}$ allows one to reduce this discrepancy!

Let us assume $f_e \sim f_q$ (although it may be quite challenging to build a theory that predicts $f_{\nu} < f_e \sim f_q$).
On one hand, the NuTeV anomaly can be explained by taking $m_{\zp}<$ GeV. On the other hand,  
predictions for $e^+ e^- \rightarrow e^+ e^-$ and $e^+ e^- \rightarrow q \bar{q}$ cross sections will now differ from the 
SM expectations by a few percent.  
These two cross sections have been well measured at LEP II. The intriguing point is that small deviations have been found in a 
preliminary analysis \cite{lepii}. It is probably too soon to conclude that these deviations are i) significant, ii) indeed due to new physics. 
However, if they turn out to be confirmed,
the effect of a light gauge boson on these cross-sections would become of crucial interest for the astro/particle/astroparticle physics communities. 

\vs
However, since a QCD explanation is possible, one can do a test that does not involve quarks but only leptons 
(so as to get rid of the QCD corrections). 
It is based on an already proposed experiment which aims to measure the ratio
$R_{lept}=\sigma_{\nu_{\mu} e}/(\sigma_{\nu_{e} e}+\sigma_{\bar{\nu_{\mu}} e})$, where 
$\sigma_{\nu_{\mu} e}$, $\sigma_{\nu_{e} e}$ and $\sigma_{\bar{\nu_{\mu}} e}$ are the
muon neutrino-, electron neutrino- and muon anti neutrino-electron elastic scattering cross-sections, respectively  \cite{Imlay}.
If a deviation is measured (as shown in Figure~3 on the right) then it is likely to be due to the presence of a light $\zp$. 
If not, this may exclude it or, alternatively, impose even stricter limits on the $\zp$ couplings.
This test is independent on $m_{\zp}$ or universality assumptions; it probes the maximal value for $[f_e f_{\nu}]^{max}$.
If a deviation is found, one has to refer to the NuTeV findings (and evaluate the size of the QCD corrections) to get an estimate of $m_{\zp}$. 
%
%

\section*{Conclusion}
Our study indicates that the LDM scenario could have already shown up in astrophysical and particle physics processes.  
Direct evidences are needed, but the presence of heavy fermionic particles $F$ may have been already detected though the 
anomalous value of the electron $g-2$. As a major consequence, the value of the fine structure constant quoted in the CODATA and used for theoretical 
estimates of various particle physics processes  could be wrong. Instead $\alpha$ could be close to its 
``direct'' experimental value! Quantities like the theoretical estimate of the muon $g-2$ (subject of great debate since a few years), 
may have to be computed again. The existence of both $F$ and $\zp$ (if one relaxes the universality assumptions) could be 
challenged in high energy colliders.
Their discovery would hint in the direction of $N=2$ supersymmetry \cite{mirror}. 

\vs
\textbf{Acknowledgement}
The authors would like to thank I. de la Calle for his PAW expertise and M. Langer for interesting discussions.


\begin{thebibliography}{99}


\bibitem{wmap}
C.~L.~Bennett {\it et al.},
``First Year Wilkinson Microwave Anisotropy Probe (WMAP)'',  
arXiv:astro-ph/0302207.

\bibitem{bens}
C.~Boehm, T.~A.~Ensslin and J.~Silk,
J.\ Phys.\ G {\bf 30} (2004) 279
[arXiv:astro-ph/0208458].

\bibitem{bf}
C.~Boehm and P.~Fayet,
Nucl.\ Phys.\ B {\bf 683}, 219 (2004)
[arXiv:hep-ph/0305261].



\bibitem{integral}
J.~Knodlseder, {\it et.al.},
Accepted for publication in A\&A,
arXiv:astro-ph/0309442;
P.~Jean {\it et al.},
arXiv:astro-ph/0309484;



\bibitem{previous}
D.~D.~Dixon {\it et al.},
arXiv:astro-ph/9703042;
P.~A.~Milne, J.~D.~Kurfess, R.~L.~Kinzer and M.~D.~Leising,
arXiv:astro-ph/0106157.

\bibitem{pjean}
G.~Weidenspointner {\it et al.},
arXiv:astro-ph/0406178, and references therein. 


\bibitem{Prantzos}
N.~Prantzos,
arXiv:astro-ph/0404501.


\bibitem{Bertone}
G.~Bertone, A.~Kusenko, S.~Palomares-Ruiz, S.~Pascoli and D.~Semikoz,
arXiv:astro-ph/0405005.


\bibitem{511}
C.~Boehm, D.~Hooper, J.~Silk and M.~Casse, J.Paul 
Phys.\ Rev.\ Lett.\  {\bf 92}, 101301 (2004)
[arXiv:astro-ph/0309686].


\bibitem{nfw}
J.~F.~Navarro, C.~S.~Frenk and S.~D.~White,
Astrophys.\ J.\  490, 493 (1997).


\bibitem{Hooper}
D.~Hooper and L.~T.~Wang,
arXiv:hep-ph/0402220;
C.~Picciotto and M.~Pospelov,
arXiv:hep-ph/0402178.


\bibitem{llbfs}
P.~Fayet,
arXiv:hep-ph/0403226.


\bibitem{Marciano}
W.~J.~Marciano,
Int.\ J.\ Mod.\ Phys.\ A {\bf 19S1} (2004) 77, and references 
therein.

\bibitem{Mohr}
P.~J.~Mohr and B.~N.~Taylor,
Rev.\ Mod.\ Phys.\  {\bf 72}, 351 (2000).




\bibitem{E821}
G.~W.~Bennett {\it et al.}  [Muon g-2 Collaboration],
Phys.\ Rev.\ Lett.\  {\bf 92}, 161802 (2004)
[arXiv:hep-ex/0401008].


\bibitem{charmii}
P.~Vilain {\it et al.}  [CHARM-II Collaboration],
Phys.\ Lett.\ B {\bf 335}, 246 (1994).


\bibitem{moi}
C.~Boehm,
arXiv:hep-ph/0405240;
S.~Davidson, S.~Forte, P.~Gambino, N.~Rius and A.~Strumia,
JHEP {\bf 0202} (2002) 037
[arXiv:hep-ph/0112302].


\bibitem{nutev}
K.~S.~McFarland and S.~O.~Moch,
arXiv:hep-ph/0306052.



\bibitem{Kretzer}
S.~Kretzer, F.~Olness, J.~Pumplin, D.~Stump, W.~K.~Tung and M.~H.~Reno,
arXiv:hep-ph/0312322.


\bibitem{lepii}
  [LEP Collaborations],
arXiv:hep-ex/0212036.




\bibitem{Imlay}
R.~Imlay and G.~J.~VanDalen,
J.\ Phys.\ G {\bf 29}, 2647 (2003).


\bibitem{mirror} 
P. Fayet, \nucl{149}{137}{1979}; Proc. 1980 Karpacz 
School (Harwood, 1981) p.115; \physl{142}{263}{1984}; 
\nucl{246}{89}{1984};
Nucl.\ Phys.\ B {\bf 187} (1981) 184;
Phys.\ Lett.\ B {\bf 95} (1980) 285.


\end{thebibliography}
\end{document}